# Magnetic Skyrmions in Atomic Thin $CrI_3$ Monolayer

Aroop K. Behera[1], Sugata Chowdhury[2], Suprem R. Das[1,3,*]

[1] Industrial and Manufacturing Systems Engineering, Kansas State University, Manhattan, Kansas 66503, USA

[2] Department of Chemical and Biomolecular Engineering, Rice University, Houston, Texas 77251, USA

[3] Electrical and Computer Engineering, Kansas State University, Manhattan, Kansas 66503, USA

**Abstract**

In this letter, we report the visualization of topologically protected spin textures, in the form of magnetic skyrmions, in recently discovered monoatomic-thin two-dimensional $CrI_3$. By combining density functional theory and atomistic spin dynamic simulation, we demonstrate that an application of out-of-plane electric field to $CrI_3$ lattice favors the formation of sub-10 nm skyrmions at 0 K temperature. The spin texture arises due to a strong correlation between magneto-crystalline anisotropy, Dzyaloshinskii-Moriya interaction and the vertical electric field, whose shape and size could be tuned with the magnetic field. Such finding will open new avenues for atomic-scale quantum engineering and precision sensing.

***Corresponding author:** Suprem R. Das (srdas@ksu.edu)

With Moore's law of device scaling seeking transformative changes for future of electronics, there has been proposals with number of alternate ideas among which spin-based electronics (spintronics) constitutes a paradigm shift.[1] Electronic spin angular momentum (or simply, spin), a fundamental degree of freedom, could be coupled and manipulated in the periodic crystals in solids in number of ways that makes the spintronics one of the intriguing researches in condensed matter physics and device engineering. Moreover, since the discovery of graphene in 2004 there has been a surge of two-dimensional (2D) materials with wide variety of electronic properties such as semimetals (e.g., graphene), semiconductors (e.g., $MoS_2$), metals (e.g., $TaS_2$), superconductors (e.g., $NbSe_2$), and insulators (e.g., h-BN).[2-4] Spin-lattice coupling is a growing field of interest in these emerging nanoscale materials with the ultimate goal to study the spin transport to develop 2D spintronics. For example, graphene has been shown to have high electrical spin transport at room temperatures and molybdenum disulfide ($MoS_2$) has been shown to have high optical spin/valley polarization.[5,6] Magnetic skyrmions, a quasiparticle with topologically spin textures hosted in condensed matter system, are of great recent interest due to both fundamental physics study of spins as well as their applications in logic and memory devices.[7,8] Since its first theoretical proposal in 2006 in condensed matter systems by Robler et al. owing to the chiral interactions led by lack of inversion symmetry, there has been number of material systems, both in bulk and thin films, that demonstrate this spontaneous ground states without the assistance of external field and/or the proliferation of topological defects.[9,10,7] While bulk materials such as MnSi show a Bloch type skyrmions with chiral interactions and/or topological defects, a Néel type skyrmions are generally observed in thin films and surfaces with broken inversion symmetry at the interface. The above two growing fields in nanoscale materials (2D materials and Néel type skyrmions) were obscured to imagine as until recently 2D-magnets were not realized experimentally and was theoretically challenged by Mermin and Wagner.[11] However, after Huang et al. and Gong et al. demonstrated the first atomic thin 2D magnets recently there have been new interest in possible observation of 2D skyrmions.[12-16] $CrI_3$, with an Ising-type ferromagnetic ordering in monolayer at temperature below 45 K, shows a crossover from ferromagnetic to antiferromagnetic in the monolayer-bilayer transition. $CrI_3$ bilayer demonstrates electrical field control of 2D linear magnetoelectric-induced antiferromagnetic-ferromagnetic switching.[17] Both Dzyaloshinskii-Moriya type interaction (DMI) (due to broken inversion symmetry and spin-orbit coupling) induced skyrmions as well as 2D Moire (due to van der Waals heterostructure of ferromagnetic monolayer on antiferromagnetic substrate) induced skyrmions have been recently proposed.[15,16] Although the theoretical report by Liu et al. has shown very important finding for future DMI-induced discovery of experimental 2D skyrmions, understanding on their large scale

(sample scale) evolution and dynamics is missing that is further needed for skyrmionics device physics study. The study, furthermore, has been made on density functional theory (DFT) calculations, which only analyzes few atoms, limiting a device scale geometry to study the skyrmionic dynamics. This present letter, by exploiting a combined DFT and an atomistic spin dynamics simulation, reports on the DMI-induced skyrmionic study, particularly their spatial-temporal evolution from an initial randomly spin-polarized state as well as their size dependent study on magnetic field.

**DFT calculation:** Calculations were carried out using DFT[18, 19] as implemented in QUANTUM ESPRESSO code.[20] We used the PBEsol generalized gradient approximation (GGA) as exchange and correlation potential.[21] We have used fully relativistic norm-conserving pseudopotentials.[22, 23] A plane-wave cutoff of 70 Ry, a 10x10x6 Monkhorst-Pack[24] k-point mesh for bulk geometries and a $8 \times 8 \times 1$ k-point grid for slab geometries were used. For calculations with Cr, we used DFT+U[25,26] with U=3 eV,. All the geometric structures are fully relaxed until the force on each atom is less than 0.002 eV/Å, and the energy-convergence criterion was 1x10$^{-6}$eV. Results from our DFT calculations were then used as input to construct maximally localized Wannier functions using WANNIER90.[29, 28]

**Spin Dynamics Simulation:** Although the present letter extensively reports the spin dynamics simulation to visually represent the skyrmionics structure in CrI$_3$, first the electric-field induced magneto-crystalline anisotropy energy (MAE, K) and the DMI interaction were calculated following density functional theory (DFT) and are later used as input parameters for the spin dynamics simulation. Both the DFT and spin dynamics simulation (for studying the dynamic behavior of the magnetic moments) were made in a monolayer CrI$_3$ configuration. Bulk CrI$_3$ crystallizes in a rhombohedral ($R\bar{3}$) symmetry below T~ 210 K and a monoclinic ($C2m$) crystal symmetry above it.[29] However, irrespective of the two symmetries, their monolayers become identical due to their relative difference in the *c* direction [see Fig.1]. In a CrI$_3$ monolayer the magnetic moments are present on the Cr atoms which form a graphene like (honeycomb) hexagonal lattice. Therefore, a continuous magnetic vector field approximation used in micromagnetic approach does not consider the underlying honeycomb lattice and could not provide an accurate spin dynamics calculation. Atomistic spin dynamics (ASD) simulation as used in this study, on other hand, considers the underlying crystal lattice and thus could be used to construct each spin vector arising because of magnetic moment on every chromium atom.[30,31] Recently developed Spirit ASD simulation tool was used for CrI$_3$ monolayer structure for the solution of Landau-Lifshitz Gilbert (LLG) equation [30]

$$\frac{d\boldsymbol{m}}{dt} = -|\gamma|\boldsymbol{m} \times \boldsymbol{H}_{eff} + \alpha\left(\boldsymbol{m} \times \frac{d\boldsymbol{m}}{dt}\right) \qquad [1]$$

Where, $\gamma$ is the gyromagnetic ratio, $m$ is the magnetic moment vector of each chromium atom, $\alpha$ is the Gilbert damping coefficient and $H_{eff}$ is the effective magnetic field of the system given by

$$H_{eff}(x) = -\nabla H(x) \qquad [2]$$

Where, $H$ is the Hamiltonian of the system.[31] The Hamiltonian consists of exchange, anisotropy, Zeeman and Dzyaloshinskii-Moriya interaction (DMI) terms. Hence the Hamiltonian can be written as:

$$H = H_{ex} + H_{ani} + H_z + H_{DMI} \qquad [3]$$

where $H_{ex}$, $H_{ani}$, $H_z$ and $H_{DMI}$ are the individual Hamiltonian terms for exchange, anisotropy, Zeeman and DMI respectively. Upon expansion, the Hamiltonian becomes

$$H = -\sum_{<ij>} J_{ij} \mathbf{n}_i \cdot \mathbf{n}_j - \sum_i \sum_j K_j (\hat{K}_j \cdot \mathbf{n}_i)^2 - \sum_i \mu_i \mathbf{B} \cdot \mathbf{n}_i - \sum_{<ij>} \mathbf{D}_{ij} \cdot (\mathbf{n}_i \times \mathbf{n}_j) \qquad [4]$$

Where, $J_{ij}$ is the Heisenberg symmetric exchange, $K_j$ is the single ion magnetic anisotropy, $\mathbf{B}$ is the external magnetic field and $D_{ij}$ is the DMI. The pairing index denotes the unique set of interacting spins at the respective sites. $\hat{K}_j$ denotes the direction of the anisotropy and the magnetic moment is described as $\mathbf{m}_i = \mu_i \mathbf{n}_i$.

The simulation set up was made by placing two spins in a sublattice/unit cell formed by a graphene-like hexagonal symmetry honeycomb lattice of chromium atoms. This spin-lattice mimics the magnetic moments present on the Cr atoms in a $CrI_3$ monolayer. We have considered a system of fifty unit-cells along $a$ and $b$ directions (see Figures). Recent works.[14,32,33] on a monolayer $CrI_3$ suggested that the magnetic anisotropy and the DMI can be tuned in a monolayer $CrI_3$ by the application of an electric field in out of plane direction. At equilibrium conditions (no external perturbation applied to the system), a $CrI_3$ lattice possesses an inversion crystal symmetry. While an applied electric field in the vertical direction to the plane of $CrI_3$ causes both the iodine planes to move in opposite directions it will move the chromium plane along the direction of the field, leading to a breaking of the inversion symmetry. This breaking of inversion symmetry with unequal-distanced iodine planes from chromium plane leads to Rashba-type the spin-orbit coupling giving rise to a net DMI in the system. We have calculated both the magneto-crystalline anisotropic energy, MAE (K) and the DMI values of 0.51 meV and 0.18 meV respectively, per Cr atom, with vertical electric field of 2.0 V/nm using DFT calculation. The out-of-plane electric field, however, had a little effect on the exchange energy and the magnetic moments associated with the Cr-I bonds, consistent with report by Liu et al.[14] For the spin dynamics calculation, we have considered the Néel type DMI corresponding to different electric field values. Periodic boundary conditions were considered in our system to rule out any possible effects arising from shape-induced anisotropy. For the spin dynamics simulation, we set the CrI$_3$ system in a paramagnetic state (in contrast to its ferromagnetic ordering), where all the spins are aligned randomly using a random seed value in the simulation and let the system evolve with the dynamics following LLG equation mentioned above. For consistency, the initial spin states and the random seed value were both kept uniform in all our simulations carried out for various cases. All our simulations were carried out in an idealistic temperature of zero kelvin.

Figure 1a and 1b show the side view and top view crystal structure of multilayer CrI$_3$ in Rhombohedral ($R\bar{3}$) symmetry respectively. Figure 1c and 1d show the side view and top view crystal structure of CrI$_3$ multilayer, respectively, but in monoclinic ($C2m$) symmetry. The chromium atoms are shown in blue spheres and the iodine atoms, bonded to the top and bottom of chromium atoms, are shown in purple spheres. Fig. 1a and Fig. 1c also show the relative stacking of the van der Waal's layers in both the geometries. Figure 1e and 1f show the atomic arrangements of chromium and iodine atoms in a monolayer CrI$_3$ in a side view and top view configuration respectively. It should be noted that the relative arrangement of Cr atoms in a CrI$_3$ monolayer is independent of its $R\bar{3}$ and $C2m$ crystal structure. Our DFT calculated lattice parameter values for $R\bar{3}$, $C2m$ bulk crystal structures and monolayer of CrI$_3$ that are shown in Table 1 and are consistent with the earlier reports.[14,29,34] Other crystal parameters, such as the three lattice angles, Cr-I bond length, the formation energy (ground state $R\bar{3}$ and excited state $C2m$), and the relative formation energy of monolayer-to-bilayer CrI$_3$ are mentioned in the Table 1. The magnetic moment, gyromagnetic ratio, and the Heisenberg exchange were calculated to be 3 $\mu_B$, 0.23 and 2.53 meV respectively, which are comparable to the previous reported values.[14-16, 34] Here $\mu_B$ stands for a Bohr's magneton, fundamental unit of spin angular momentum. The yellow arrows pointing all up in Figure 1g denote the magnetic moments of adjacent chromium atoms showing a ferromagnetic (FM) coupling in a monolayer CrI$_3$. DFT calculations were also employed to calculate the MAE (K) value at equilibrium conditions. However, owing to its lack of crystal inversion symmetry, CrI$_3$ does not possess a DM interaction at equilibrium conditions.

Figure 2a shows the calculated band structure of monolayer CrI$_3$ along the crystallographic points shown in horizontal direction with energy in vertical direction. A band gap of ≈ 0.8 eV at Γ-point is in good agreement with previously published report.[35] Next, using the non-collinear self-consistent field DFT method, we compute the important parameter MAE (K) under an vertical external applied electric field of 2.0 V/nm (as we will discuss later), where it becomes maximum at $\theta \approx 90^0$ (where $\theta$ is the angle between Cr-I bond and the in-plane CrI$_3$ structure). The angle dependent MAE (K) values of monolayer CrI$_3$ is shown in Figure 2b (the inset shows the optimized crystal structure).

The sample for atomistic spin dynamics simulation was constructed by using a spin lattice of CrI$_3$ (a lattice with magnetic moments of Cr atoms, arranged in a honeycomb graphene-like geometry), as shown in Figure 3a. All the spins shown in white are aligned in $+z$ direction. The colour map for spins is shown in lower right corner of Figure 3a, depicting an RGB pattern. By beginning our simulation from a paramagnetic state and using the equilibrium magneto-crystalline anisotropy and zero DMI of CrI$_3$, no spin texture was visible as expected. Later, an external electric field perturbation to the CrI$_3$ structure was implemented in the simulation by using DFT calculated DMI, K, and J (exchange) values (similar values are also reported in ref. 14, 15). Even with an electric field of $|\mathbf{E}| = 1.2$ V/nm (with 0 T out-of-plane magnetic field), a saturated FM state was obtained as the stable configuration, similar to the one shown in Figure 3a. This could be understood from the underlying competitive interactions as governed by the Hamiltonian (equation 4): while the Zeeman energy will cause topologically unaffected to the spin structure, the exchange and the anisotropy terms tend to align the spins in a collinear manner but the DMI tend to rotate the spins away from each other (DMI will instead cause a canted configuration.). With the exchange and anisotropy apparently dominating over the DMI in the case of $|\mathbf{E}| = 1.2$ V/nm, a FM single domain state is favored causing to see a saturated spin structure in Figure 3a. Note that a system having just DMI term in the Hamiltonian will always experience a spiral state as the ground state. Therefore, a spin saturation (FM ground state) in Figure 3a indicates a smaller DMI caused by the electric field that is unable to spiral the spins. To validate our arguments, we considered a hypothetical case where we kept all other parameters same as the above case ($|\mathbf{E}| = 1.2$ V/nm field perturbation) but reduced the anisotropy energy per Cr atom by ~ 70%. Subsequently the spin dynamics simulation was performed on the system with same initial paramagnetic state and the system was allowed to relax in time to attain the ground state at 0 T magnetic field. It is important to mention that, although a ~70% reduced MAE (K) case does not correspond to any applied electric field strengths in our DFT calculation but such an effect and quantification in the reduction of MAE with external field is of future interest to us. The ASD simulation leads to a mixture of chiral domains, skyrmions and skyrmions with opposite chirality mediated through a 540° domain walls as seen in other counterparts such as thin film magnets.[36] The simulation was further conducted by keeping the parameters unchanged but by varying the magnetic field from 0 T to 1.0 T. Figure 3b to Figure 3e shows that the size of skyrmions gradually reduces and the density of chiral domains in the CrI$_3$ sample decreases with the increase in magnetic field. Figure 3f shows the magnified version of one of the skyrmions at the stable ground state achieved at 1.0 T magnetic field (Figure 3e).

Number of studies on different 2D materials are being reported on the application of vertical electric field to understand the effects of crystal structure in monolayer limit, and its correlation with strain and band gap etc.[37] A band gap tuning of 0 to 250 meV in bilayer graphene and it can be substantially tuned in rippled monolayer MoS$_2$ by an application of vertical electric fields, leading to improved device performances.[38,39] Application of a vertical electric field application is also carried out in case of monolayer CrI$_3$, where Liu et al. have shown a tuning of MAE and DMI with the application of the external field.[14,15] Our DFT calculations for equilibrium CrI$_3$ monolayer structure as well as a field of 2.0 V/nm is consistent with the report by Liu et al. With a $|\mathbf{E}| = 2.0$ V/nm applied field and considering a Néel type DMI (in-plane of CrI$_3$), we observe a 60 % increase in the DMI energy and a 28 % decrease in the anisotropy energy from its values calculated at $|\mathbf{E}| = 1.2$ V/nm. Such an effect is significant in CrI$_3$ due to the FM ground state coupling. As a result of the DMI, the system breaks the inversion symmetry leading to an increase in spin-orbit coupling (Rashba-type). On the other hand, the opposing trends in the DMI and the MAE values favors the spin spiraling effects we discussed earlier. ASD simulation was then performed on the equilibrium structure with a paramagnetic ground state and with the new DMI and K input parameters. As shown in Figure 4a, even at a zero magnetic field, a $|\mathbf{E}| = 2.0$ V/nm applied vertical electric field to CrI$_3$ monolayer lattice generates chiral domains and onset of skyrmions. Subsequent simulations were carried out for increasing out-of-plane magnetic fields and the stable ground states were observed. Figure 4b to Figure 4d show gradual decrease in the chiral domain density and formation and gradual increase of more skyrmionic states in the system. As discussed earlier in the hypothetical case, while the exchange and anisotropy aligns the spins along the FM ground state the Rashba spin-orbit induced DMI is strong enough to spiral them away, eventually forming the topological spin textures. DMI is strong enough to cause these competing effects even at zero magnetic field. As earlier, Figure 4e shows the magnified version of one of the skyrmions at the stable ground state achieved at 1.0 T magnetic field. The skyrmion diameter (D, as indicated in Figure 4e) was calculated by counting the number of lattice constants over which the chromium spin flips from an up-state at the periphery to down-state at the center and then again to up-state at the opposite periphery. A consistent decrease in size of the skyrmions was observed with increasing the magnetic field, with maximum size of 6.9 nm at zero magnetic field and minimum of 4.14 nm at 1.0 T field. Figure 4f shows the plot of the size of skyrmions vs. the magnetic field for a fixed electric field of 2.0 V/nm.

In summary, although a monolayer of CrI$_3$ lacks breaking of inversion symmetry, a vertical electric field of 2 V/nm induces appropriate magneto-crystalline anisotropy and DMI that combine favors hosting of skyrmions of sub-10 nm diameters at 0 K temperature. The spin textures evolve with chiral domains at zero magnetic fields to fully formed skyrmions at about 1T. Moreover, they show characteristic features of decreasing their size with increasing magnetic field, similar to their earlier observation in thin film magnets interfaced with metals with high spin-orbit coupling. Our work will provide guidelines for experimental observation of atomic thin skyrmions in monolayer magnets and paves the way for use of van der Waals magnets for atomic-scale quantum engineering and precision sensing.

**Acknowledgements**

S.R.D. acknowledges funding from the Department of Industrial and Manufacturing Systems Engineering, and Carl R. Ice College of Engineering, Kansas State University, Manhattan, Kansas. S.C. acknowledges National Institute of Standards and Technology (NIST), Gaithersburg, Maryland, for computational support. S.C. and S.R.D. thank Dr. Curt Richter, Dr. Angela R. Hight Walker, Dr. Carelyn E. Campbell, and Dr. Francesca Tavazza at NIST for helpful discussion. S.R.D. and A.K.B. thank Gideon Muller and Nikolai Kiselev at Forschungszentrum Jülich for helpful discussion.

**References**

[1] S. Manipatruni, D.E. Nikonov and I. A. Young, Nature Phys. **14**, 338 (2018).

[2] B. Radisavljevic, A. Radenovic, J. Brivio, V. Giacometti and A. Kis, Nature Nanotechnology 6, 147 (2011).

[3] K. S. Novoselov, D. Jiang, F. Schedin, T. J. Booth, V. V. Khotkevich, S. V. Morozov, and A. K. Geim, PNAS 102, 10451 (2005).

[4] V. K. Sangwan and M. C. Hersam, Annu. Rev. Phys. Chem. 69, 299 (2018).

[5] N. Tombros, C. Jozsa, M. Popinciuc, H. T. Jonkman & B. J. van Wees, Nature 448, 571 (2007).

[6] Y. K. Luo, J. Xu, T. Zhu, G. Wu, E. J. McCormick, W. Zhan, M. R. Neupane, and R. K. Kawakami, Nano Lett. 17, 3877 (2017).

[7] A. Fert, V. Cros, and J. Sampaio, Nature Nanotechnology 8, 152 (2013).

[8] K. Everschor-Sitte, J. Masell, R. M. Reeve and M. Kläui, J. Appl. Phys. 124, 240901 (2018)

[9] U. K. Roßler, A. N. Bogdanov, and C. Pfleiderer, Nature 442, 797 (2006).

[10] S. Mühlbauer, B. Binz, F. Jonietz, C. Pfleiderer, A. Rosch, A. Neubauer, R. Georgii, and P. Böni, Science 323, 915 (2009).

[11] N.D. Mermin, and H. Wagner, Phys. Rev. Lett. 17, 1133 (1966)

[12] B. Huang, G. Clark, E. Navarro-Moratalla, D. R. Klein, R. Cheng, K. L. Seyler, D. Zhong, E. Schmidgall, M. A. McGuire, D. H. Cobden, W. Yao, D. Xiao, P. Jarillo-Herrero, and X. Xu, Nature 546, 270 (2017)

[13] C. Gong, L. Li, Z. Li, H. Ji, A. Stern, Y. Xia, T. Cao, W. Bao, C. Wang, Y. Wang, Z. Q. Qiu, R. J. Cava, S. G. Louie, J. Xia, and X. Zhang, Nature 546, 265 (2017)

[14] J. Liu, M. Shi, J. Lu, and M. P. Anantram, Phys. Rev. B97, 054416 (2018).

[15] J. Liu, M. Shi, P. Mo, and J. Lu, AIP Adv. **8**, 055316 (2018).

[16] Q. Tong, F. Liu, J. Xiao, and W. Yao, Nano Lett. 18, 7194 (2018).

[17] S. Jiang, J. Shan, and K. F. Mak, Nature Materials 17, 406 (2018).

[18] P. Hohenberg and W. Kohn, Physical review **136** (3B), B864 (1964).

[19] W. Kohn and L. J. Sham, Physical review **140** (4A), A1133 (1965).

[20] P. Giannozzi, S. Baroni, N. Bonini, M. Calandra, R. Car, C. Cavazzoni, D. Ceresoli, G. L. Chiarotti, M. Cococcioni and I. Dabo, Journal of physics: Condensed matter **21**, 395502 (2009).

[21] J. P. Perdew, A. Ruzsinszky, G. I. Csonka, O. A. Vydrov, G. E. Scuseria, L. A. Constantin, X. Zhou and K. Burke, Physical Review Letters **100**, 136406 (2008).

[22] D. Hamann, Physical Review B **88**, 085117 (2013).

[23] M. Schlipf and F. Gygi, Computer Physics Communications **196**, 36-44 (2015).

[24] H. J. Monkhorst and J. D. Pack, Physical review B **13**, 5188 (1976).

[25] A. Liechtenstein, V. Anisimov and J. Zaanen, Physical Review B **52**, R5467 (1995).

[26] S. Dudarev, G. Botton, S. Savrasov, C. Humphreys and A. Sutton, Physical Review B **57**, 1505 (1998).

[27] N. Marzari and D. Vanderbilt, Physical review B **56**, 12847 (1997).

[28] A. A. Mostofi, J. R. Yates, Y.-S. Lee, I. Souza, D. Vanderbilt and N. Marzari, Computer physics communications **178**, 685-699 (2008).

[29] M.A. McGuire, H. Dixit, V. R. Cooper, and B. C. Sales, Chem. Mater., 27, 612 (2015).

[30] G. P. Muller, M. Hoffmann, C. Disselkamp, D. Schurhoff, S. Mavros, M. Sallermann, N. S. Kiselev, H. Jonsson, and S. Blugel, arXiv:1901.11350 [cond-mat.mes-hall] (2019).

[31] J.H. Mentink, M.V. Tretyakov, A. Fasolino, M. I. Katsnelson, and T. Rasing, J. Phys.: Condens. Matter 22, 176001 (2010).

[32] L. Webster, and J. -A. Yan, Phys. Rev. B98, 144411 (2018).

[33] J. L. Lado, and J. Fernandez-Rossier, 2D Mater. 4, 035002 (2017)

[34] W. -B. Zhang, Q. Qu, P. Zhu, and C. -H. Lam, J. Mater. Chem. C, 3, 12457 (2015).

[35] P. Jiang, L. Li, Z. Liao, Y. X. Zhao, and Z. Zhong, Nano Lett. 18, 3844 (2018).

[36] C. -Y. You, and N. -H. Kim, Current Appl. Phys. 15, 298 (2015).

[37] M. Sharma, A. Kumar, P.K. Ahluwalia, and R. Pandey, J. Appl. Phys. 116, 063711 (2014).

[38] Y. Zhang, T. -T. Tang, C. Girit, Z. Hao, M. C. Martin, A. Zettl, M. F. Crommie, Y. R. Shen, and F. Wang, Nature 459, 820 (2009)

[39] J. Qi, X. Li, X. Qian, and J. Feng, Appl. Phys. Lett. 102, 173112 (2013)

**Table: Calculated crystal parameters, formation energy and monolayer-bilayer formation energy of $CrI_3$.**

| Lattice Parameters | $\bar{R}3$ | C2/m |
|---|---|---|
| a (Å) | 6.91 | 6.89 |
| b (Å) | 6.91 | 11.79 |
| c (Å) | 19.83 | 6.93 |
| α | 90 | 90 |
| β | 90 | 108.63 |
| γ | 120 | 90 |
| Cr-I | 2.88 | 2.73 |
| **Energy Difference** | | |
| $E_{formation}$ (meV/atom) | 0 | 2.1 |
| **Monolayer** | | |
|  | DFT (this work) | Experiment (ref. 29) |
| a (Å) | 6.963 | 6.867 |
| Cr-I bond length $l$ (Å) | 2.76 | 2.727 |
| **Relative formation Energy Monolayer to Bilayer:** $E_{FM}$-$E_{AFM}$ = 0.038 eV | | |

**Figure caption**

Figure 1. Crystal structure of $CrI_3$: Bulk $CrI_3$ van der Waal stacking in rhombohedral ($R\bar{3}$) symmetry (side view, in a and top view in b). (c) and (d) show side view and top view, respectively, of bulk $CrI_3$ van der Waal stacking in monoclinic ($C2m$) symmetry. Figure (e) and (f) show the side view and top view of monolayer $CrI_3$ respectively. Figure (f) shows the ferromagnetic ordering of monolayer $CrI_3$.

Figure 2. (a) Energy band structure of $CrI_3$ monolayer using DFT calculations, showing a band gap opening of 0.8 eV at Γ-point; (b) Angular dependence of magneto-crystalline anisotropic energy (MAE, K) with change of the spin orientations of Cr atoms (i.e., Cr-I chemical bond) from its horizontal alignment (i.e., in-plane direction).

Figure 3. $CrI_3$ spin lattice for atomistic spin dynamics calculation. (a) The magnetic moments of Cr atoms are arranged in a honeycomb graphene-like geometry. The white color indicates spin alignment in $+z$ direction, bottom right inset shows the spin orientation color map and top left inset shows the crystallographic axes. Figure (b) – (e) show the spin dynamics calculation (LLG) and the ground state configuration of spins at various magnetic fields when $CrI_3$ undergoes a vertical electric field 1.2 V/nm and with a hypothetical ~ 70% reduced anisotropy, forming mixture of chiral domains and skyrmions to pure skyrmionic states. Figure (f) shows a magnified view.

Figure 4. Atomistic spin dynamics calculation for $CrI_3$ spins with actual external electric field of 2.0 V/nm applied vertically to the plane of $CrI_3$. DFT calculated parameters (such as the DMI and K values at 2.0 V/nm) are used as the spin dynamics simulation inputs. Figure (a) – (d) show the spin dynamics calculation (LLG) and the ground state configuration of spins at various magnetic fields, forming mixture of chiral domains and skyrmions to pure skyrmionic states. Figure (e) shows the magnified image and the diameter of a single skyrmion. Figure (f) shows a gradual decrease of skyrmionic diameter with applied magnetic field.

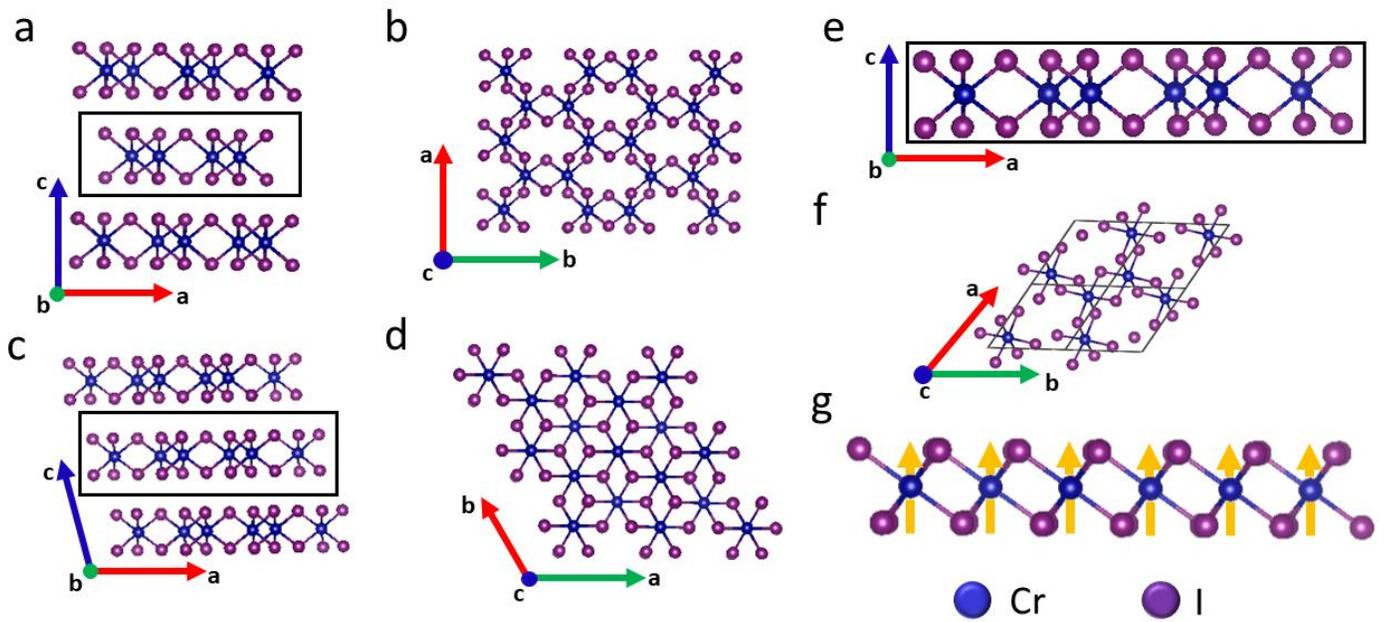

Figure 1 *Behera et al.*

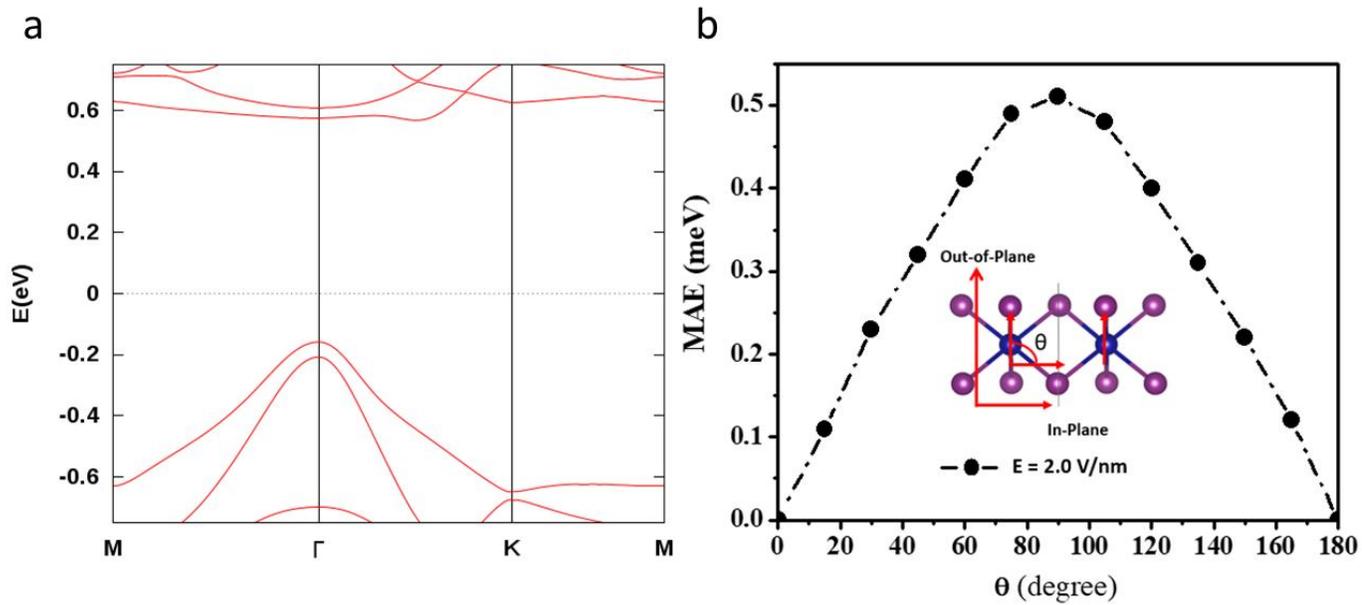

Figure 2 *Behera et al.*

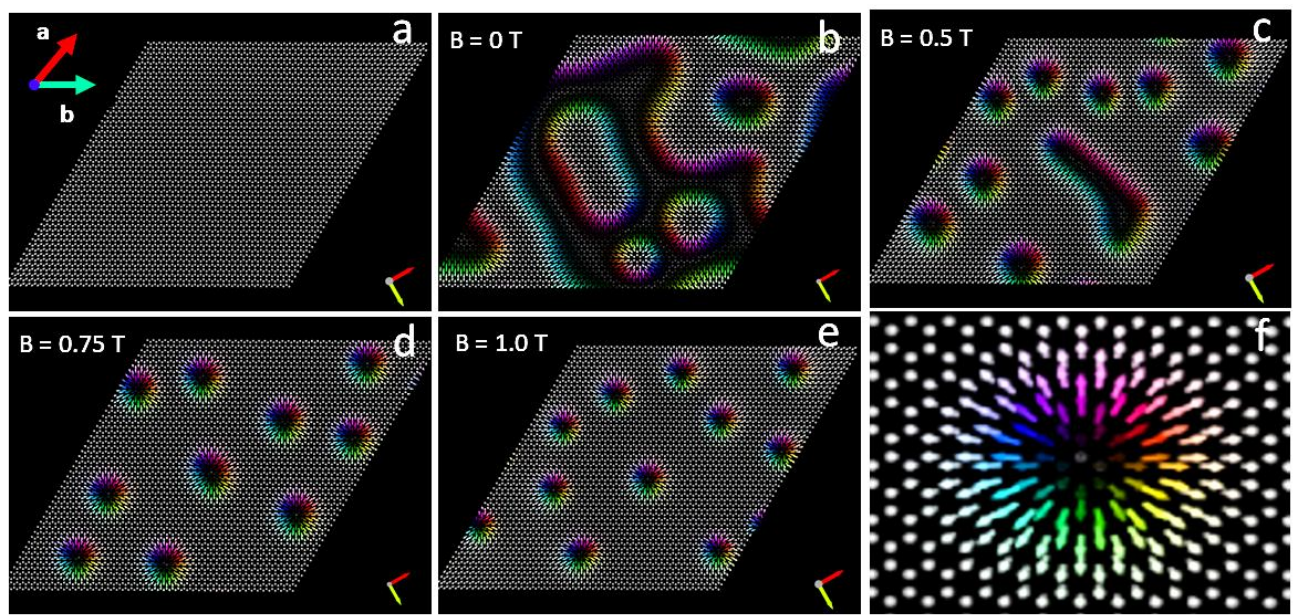

Figure 3 *Behera et al.*

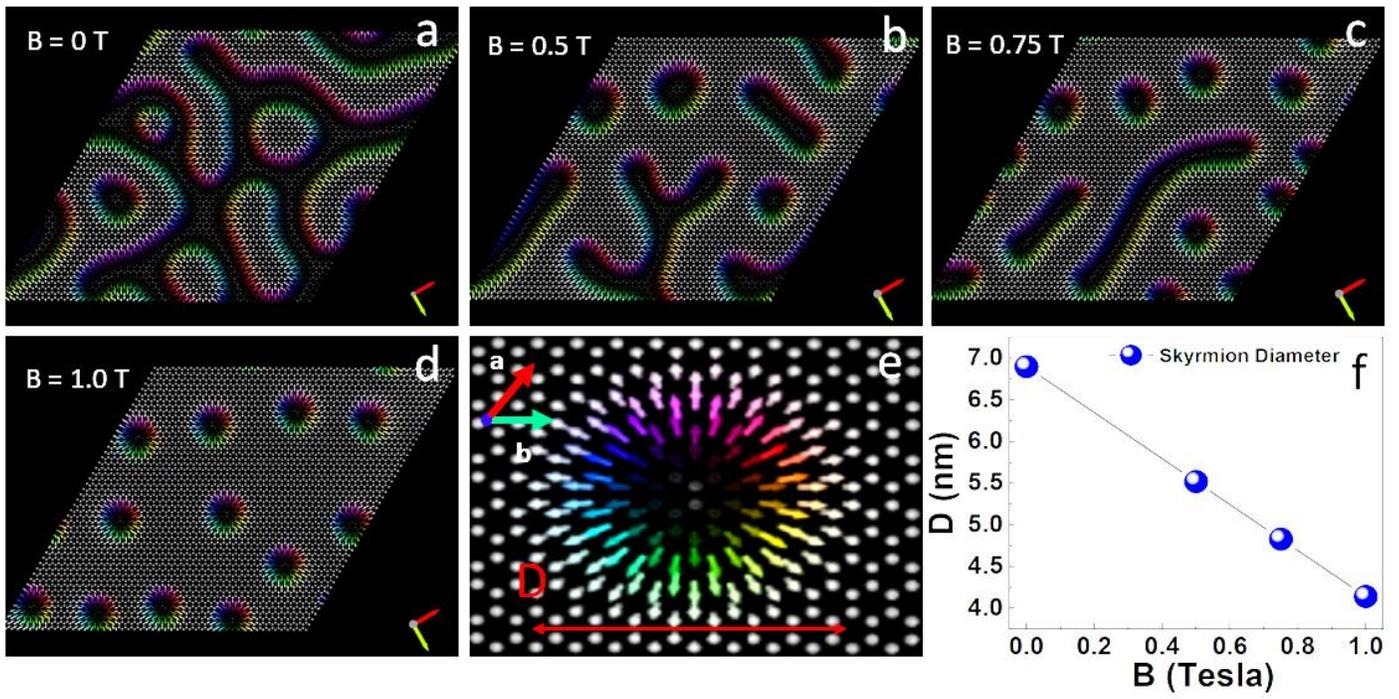

Figure 4 *Behera et al.*